\documentclass[12pt]{article}

\usepackage{graphics,cite,amssymb,epsfig,float,psfrag}
\usepackage[usenames,dvips]{color}
\usepackage{rotating}
\newcommand{\lsim}{\raisebox{-0.13cm}{~\shortstack{$<$ \\[-0.07cm] $\sim$}}~} 
\newcommand{\gsim}{\raisebox{-0.13cm}{~\shortstack{$>$ \\[-0.07cm] $\sim$}}~} 
\newcommand{\beq}{\begin{eqnarray}} 
\newcommand{\eeq}{\end{eqnarray}}

\begin{document}

\vspace{.8cm}

\hfill  LPT--ORSAY--12--87


\vspace*{1.4cm}

\begin{center}

{\large\bf Precision Higgs coupling measurements at the LHC}
 
\vspace*{2mm} 

{\large\bf through ratios of production cross sections}

\vspace*{.6cm}

{\sc Abdelhak Djouadi} 

\vspace*{.8cm}

\begin{small}

Laboratoire de Physique Th\'eorique, U. Paris--Sud and  CNRS,  F--91405
Orsay, France.

\vspace*{1mm}

Theory Unit, Department of Physics, CERN, CH-1211 Geneva 23, 
Switzerland.

\end{small}

\end{center}

\vspace*{1cm}

\begin{abstract} 

Now that the Higgs particle has been observed by the ATLAS and CMS experiments
at the LHC, the next endeavour would be to probe its fundamental properties and
to measure its couplings to fermions and gauge bosons with the highest  possible
accuracy. However,  the measurements will be limited by significant theoretical
uncertainties that affect the production cross section in the main production
channels as well as by experimental systematical errors. Following earlier work,
we propose in this paper to consider  ratios of Higgs production cross  sections
times decay branching ratios in which most of the theoretical uncertainties and
some systematical errors, such as the ones due to the luminosity measurement and
the  Higgs decay branching fractions,  cancel out. The couplings of  the Higgs
particle could be then probed in a way that will be mostly limited by the
statistical accuracy achievable at the LHC and accuracies at the percent level 
are foreseen for some of the ratios at the end of the LHC run. At the 
theoretical level, these ratios are also interesting as they do not involve  the
ambiguities that affect the Higgs total decay width in new physics scenarios. To
illustrate how these ratios  can be used to determine the Higgs couplings, we
perform a rough analysis of the recent  ATLAS and CMS data which shows that
there is presently no significant deviation from the  Standard Model
expectation. 
\end{abstract}

\newpage

\subsection*{1. Introduction}

It is expected since a long time that the probing of the mechanism that triggers
the breaking of the electroweak symmetry and generates the fundamental particle
masses will be, at least, a two chapters story. The first one is the search and
the observation of a spin--zero Higgs particle that will confirm  the scenario
of the  Standard Model (SM) and most of its extensions, that is,  a spontaneous
symmetry breaking by a scalar field that develops a non--zero vacuum expectation
value \cite{Higgs}. This long  chapter has just been closed by the ATLAS and CMS
collaborations with the spectacular observation of a new boson with a mass of
$\approx 125$ GeV \cite{LHC,ATLAS,CMS} and with, apparently,  the basic properties
required by the symmetry breaking mechanism in the SM \cite{Adam,Chris,Fits}.
This crucial observation opens a second and equally important chapter:  the
precise determination of the Higgs boson profile and  the unraveling of the 
mechanism itself. In particular, a precise  measurement of the Higgs couplings
to fermions and gauge bosons (as well as its self--coupling) will be mandatory
to establish the exact nature of the symmetry  breaking mechanism and,
eventually, to pin down effects of new physics if additional ingredients beyond
those of the SM are involved \cite{Review}.

Fortunately, the Higgs particle was born under a very lucky star which will 
make this second chapter rather eventful and exciting.  Indeed, the mass value
$M_H \approx 125$ GeV allows to produce the Higgs particle at the  LHC in many
redundant channels  and to detect it in a  large variety of decay modes. This
is  illustrated in Fig.~1 where, in the left-hand side, the  decay branching
fractions of the SM Higgs boson are displayed for the narrow mass range
$M_H=120$--130 GeV and it can be seen that the decay modes into  $b\bar b$,
$\tau^+\tau^-$, $WW^*$ and $ZZ^*$ final states are significant; this is also the
case for the rare but clean loop induced decays $H\to \gamma \gamma$ and  
eventually  $H\to Z\gamma$,  and even the very rare  $H\to \mu^+\mu^+$ channel, 
which should be accessible with enough data. In the right--hand side of the
figure, shown  are the production rates at the LHC of a 125 GeV SM Higgs boson
for various past, present and foreseen center of mass energies. While the by far
dominant gluon--gluon fusion mechanism $gg\to H$ has extremely large rates, the
subleading channels, i.e. the vector boson fusion (VBF) $qq \to Hqq$, the
Higgs--strahlung (HV) $q\bar q \to HV$ with $V=W,Z$ and the top quark associated
$p\bar p\to t\bar t H$ mechanisms, have cross sections which should allow a
study of the Higgs particle with $\sqrt s\gsim 14$ TeV once a large  luminosity,
$\gsim 100$ fb$^{-1}$, has been collected.

\begin{figure}[hbtp]
\begin{center}
\vspace*{-.3cm}
\hspace*{-1.5cm}
\epsfig{file=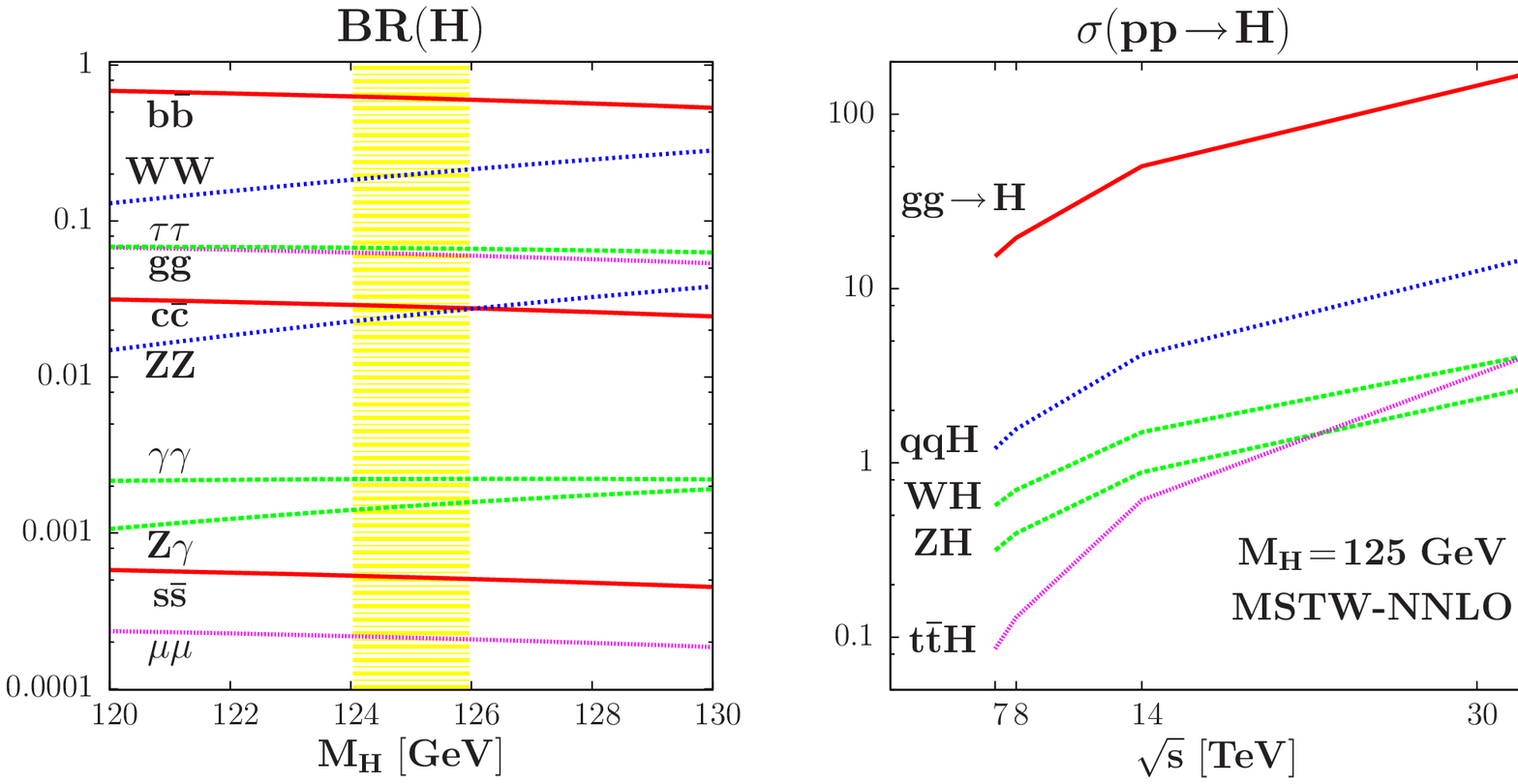,width=13.2cm}
\vspace*{-.4cm}
\caption[]{\small The branching ratios of the SM Higgs boson in the mass range 
$M_H=120$--130 GeV (left) and its production cross sections at the LHC for 
various c.m. energies (right).}
\end{center}
\label{allH}
\vspace*{-.5cm}
\end{figure}

The precision measurement  chapter is already open as, a few days only after the
Higgs  discovery, a number of theoretical analyses have appeared  to determine
the Higgs  couplings \cite{Adam,Chris,Fits}. In fact the preamble has been
written by  the ATLAS and CMS collaborations themselves \cite{ATLAS,CMS} as they
already quoted the values of the global signal strength modifier $\hat \mu$
which, with some approximation, can  be identified with the  Higgs cross section
normalised to the SM  expectation,  when  the various analyzed Higgs search
channels are combined
\begin{eqnarray}  
{\rm ATLAS}: &  \hat \mu=1.40 \pm 0.30\, \\ 
{\rm CMS}:   &  \hat \mu =0.87 \pm 0.23 \,  \label{muvalues} 
\end{eqnarray}

These first results already deliver two messages. An important first message is 
that the
observed particle seems to approximately have the couplings to fermions and
gauge bosons that are predicted by the SM, despite some small  excesses and
deficits that appear in some individual channels.  A second message is
that, already  with the rather limited statistics at hand, the accuracy of the
measurements  in eqs.~(1--2) is reaching  the 20\% level for  ATLAS and $\approx
25\%$ for CMS. This is  at the same time impressive and worrisome. Indeed,
as  illustrated in  Fig.~1, the main Higgs production channel is the top and 
bottom quark loop mediated gluon--gluon fusion mechanism, $gg \to H$, and at
$\sqrt s=7$ or 8 TeV, the  three other mechanisms contribute at a level below
15\% when their rates are added and  before kinematical cuts are applied
\cite{LHCXS,BD}. Hence, the majority of signal events presently observed at the
LHC, in particular in the main search channels  $H \to   \gamma \gamma, H  
\to    ZZ^*  \to  4\ell^\pm,  H   \to   WW^*  \to \ell^+ \nu \ell^- \bar \nu$
(with $\ell\!=\!e,\mu$) and, to a lesser extent $H   \to    \tau^+\tau^-$, come
from the $gg$ fusion mechanism which  is known to be affected by large
theoretical uncertainties. 

As a matter of fact, although the cross section $\sigma(gg\to H)$ is known up 
to next-to-next-to-leading order (NNLO) in perturbative QCD \cite{ggH-QCD}  (and
at least at NLO for the electroweak interaction \cite{ggH-EW}), there is a
significant  residual scale dependence which points to the possibility that
still higher order contributions beyond NNLO cannot be  totally excluded. In
addition, as the process is of ${\cal O}(\alpha_s^2)$ at LO and is initiated by
gluons, there are sizable uncertainties due to the  gluon parton distribution
function (PDF) and the value of the coupling $\alpha_s$. In total, the combined
theoretical uncertainty\footnote{A third source of theoretical uncertainties,
the use of an effective field theory approach to calculate the radiative
corrections  beyond the NLO approximation, should in principle also be
considered \cite{BD,BD1} and would  increase the total theoretical uncertainty
up to  $\Delta^{\rm th} \sigma(gg\to H) \approx \pm $25--30\%. In addition,
large uncertainties arise when the $gg\to H$ cross section is broken into jet
categories as will be discussed later.} has been estimated to be of  order 
$\Delta^{\rm th}  \sigma(gg\to H) \approx \pm 20\%$ by the LHC Higgs cross
section working group (LHCHWG)  \cite{LHCXS}. Hence, the theoretical uncertainty
is  already at the level of the accuracy of the measured cross section by the
ATLAS and CMS collaborations, eqs.~(1--2). 

The impact of the theoretical uncertainty can be viewed as follows
\cite{excess}. The normalisation cross section $\sigma^{\rm SM}$ adopted by the
ATLAS and CMS collaborations, and which led to the results of eqs.~(1--2), is 
borrowed from the LHCHWG. For the dominant $gg\to H$ process, it is obtained by
using the calculation performed at NNLO in QCD with a central choice for the
renormalisation and factorisation scales $\mu_F=\mu_R=\mu_0=\frac12 M_H$  (which
is approximately the same as what is obtained using the resummed cross section
at next-to-next-to-leading-logarithm with a central scale $\mu_0=M_H$)  and by
adopting  the MSTW2008  NNLO set of PDFs \cite{MSTW} to evaluate the gluon
density,  $\sigma^{\rm SM}\simeq \sigma (gg \to H)|^{\rm MSTW}_{\mu_0=\frac12
M_H}$. However, in principle, any NNLO PDF set  and any scale choice in the
conventional range $\frac12 \leq \mu/\mu_0 \le 2$ can  be adopted to evaluate
the cross section $\sigma(gg\to H)$, the difference compared  to the reference
value being accounted for by the theoretical uncertainty. For instance, if one
uses the NNPDF PDF set \cite{NNPDF} and adopts a scale choice $\mu_0=\frac14
M_H$, one would obtain a cross  section that is $\approx 20\%$ higher than
the reference cross section   
\beq
\sigma (gg \to H)|^{\rm NNPDF}_{\mu_0= \frac14 M_H} \approx 1.21  \; \sigma^{\rm 
SM}
\eeq  
In turn, if the cross section is evaluated with the ABM11 NNLO PDF set
\cite{ABM} with a scale choice $\mu_0=M_H$, one would have a cross section that
is $\approx 15\%$ lower, 
\beq
\sigma (gg \to H)|^{\rm ABM}_{\mu_0=  M_H} \approx 0.85  \; \sigma^{\rm SM}  
\eeq
Hence, if ATLAS had used the cross section value of eq.~(3) and CMS the one of
eq.~(4) for the determination of the signal strength $\hat \mu$ of eqs.~(1--2),  
they both would have had a $2\sigma$ discrepancy from the SM  expectation. In
particular, as was discussed  in Ref.~\cite{excess}, the   $\approx 2\sigma$
excess observed by the ATLAS collaboration in the $H\to \gamma\gamma$ channel 
would turn into either a simple $1\sigma$ or a  tantalising $3\sigma$ effect, 
depending on the chosen normalisation. 

It is therefore very important to eliminate this cross section normalisation
problem (or, in other words, the theoretical uncertainty) as it induces a
bias that is already now, i.e. with the $\approx 10$ fb$^{-1}$ data
presently collected by ATLAS and CMS, at the level of the experimental accuracy
of the cross section measurement. This uncertainty will be the principal
limiting factor in the extraction  of the Higgs couplings in a very near
future. 

A possibility which was considered to be very promising is the VBF production
channel, $qq \to Hqq$ with $H\to WW, \gamma \gamma$ and $\tau \tau$, in which
specific cuts reduce the various backgrounds significantly and allow a nice
extraction  of the Higgs signal \cite{VBF}. Indeed, the cross section  for the
inclusive process has been shown to have a very small combined scale and PDF
uncertainty, $\lsim \! \pm \! 3\%$ for $M_H\!=\! 125$ GeV at $\sqrt s\!=\!7$--8
or 14 TeV \cite{LHCXS}. However, it turns out that at least at $\sqrt s =7$--8
TeV, the process is contaminated by a significant fraction of gluon fusion
events,  $gg\! \to\! H\!+\!jj$, of the order of 30\% \cite{Hjj}, even  after the
specific cuts that select the VBF configuration are applied.  The  $gg\! \to\!
H\!+\!jj$ channel is affected by much larger uncertainties than  the inclusive
$gg\!\to\! H$ process, up to 50\%  when one adds the scale and the PDF
uncertainties, as well as by additional uncertainties from the jet veto when
$\sigma(gg\to H)$  is broken into jet cross sections \cite{Scale-H0j}. This
makes the total  uncertainty  in the $H\!+\!jj$ final sample that includes the
VBF part  as large as in the inclusive   $gg\to H$ case.

To remove the theoretical (and other) uncertainties,  we suggest in this paper
to simply consider ratios of production cross sections times decay branching
fractions. Similar ratios have been proposed in the past, in particular by D.
Zeppenfeld and collaborators  \cite{Dieter}, at a time when the $gg\to H\to ZZ,
WW$  channels were  not expected to be viable for Higgs masses below 130 GeV and
the focus was on the vector boson fusion processes which appeared  to be more
promising \cite{VBF}. Here, we first extend on these previous analyses by
including all channels that are expected to be observable at the LHC, in
particular those in which the $gg\! \to\! H$ cross section is broken into jet
categories and the  modes $gg\!\to\! H\! \to \! \tau^+\tau^-$ and $q\bar q\! \to
\! HV$ with boosted $H\to b\bar b$, that have not been considered in
Ref.~\cite{Dieter}. A generalized formalism for decay ratios that are free of
theory uncertainties,  denoted by  $D_{XX}$, and also cross section ratios
$C_{XX}$ which are less powerful as they still involve these uncertainties, is
introduced. For an illustration of how the proposed ratios could be used in
practice, we perform a rough analysis of the present  ATLAS and CMS data  and
conclude that there is, presently, no significant deviation of these ratios from the SM
expectation.

\subsection*{2. The decay ratios $\mathbf{D_{XX}}$}

To define ratios of Higgs production cross sections, one needs first to chose a
reference channel for the Higgs decays. The most obvious ones are the modes
$H\to VV^*$ with $V=Z,W$, which lead to the clean $ H \! \to\! ZZ^*\! \to\! 
4\ell^\pm$ or $H\! \to\!  WW^*\!  \to\!  \ell \ell \nu \nu$  final states.  One
then defines the decay ratio, that we denote by $D_{XX}$, for a given 
search   channel $H\! \to\!  XX$,  
\beq
D_{XX} = \frac {\sigma( gg \to H \to XX)}{ \sigma( gg \to H \to VV)} = 
\frac {\sigma( gg \to H)\times {\rm BR} (H \to XX)}{ \sigma( gg \to H)
\times {\rm BR} (H \to VV)} =
\frac{\Gamma( H \to XX)}{ \Gamma ( H \to VV)} \label{ratio} 
\eeq 
in which the cross section $\sigma(gg \to H)$ and hence, its significant theory 
uncertainty, cancels out\footnote{When considering the same production process,
a shift in the normalisation should  be noticed in the correlation between two
different decay channels; we thank Dieter Zeppenfeld  for a discussion on this
point.}, leaving only the ratio of decay branching fractions  and hence of
partial decay widths.  In fact, even the total Higgs decay width, which includes
the contributions of channels such as $H\to gg$ and $H\to c\bar c$ that cannot
be accessed at the LHC, as well as possible invisible Higgs decays in scenarios
beyond the SM, do  not appear in the decay ratios $D_{XX}$.  

In addition, two  other important sources of  experimental uncertainties also
cancel out:  some common experimental systematical uncertainties such as, for
instance,  the  one due to the luminosity measurement which is presently  at the
level of  a few percent, and    the uncertainties in the Higgs branching ratios
which are of the order of 3--5\% for BR($H\to b\bar b, WW,ZZ, \gamma \gamma,
\tau\tau)$. The latter uncertainties  are mainly due to the $H\to b\bar b$
partial decay width which is affected by the errors on the input values of the
bottom quark mass and the coupling $\alpha_s$ and which then migrate to the
decays branching fractions through the total Higgs decay width that is 
controlled by the $b\bar b$ mode \cite{BD,BRs-MS}. In the decay  ratios of
eq.~(\ref{ratio}), these uncertainties disappear when one considers  final
states like $H\to WW,ZZ, \gamma \gamma, \tau\tau$, where, in contrast  to  Higgs
decays into quark and gluon pairs,   only  small electroweak effects are
involved and no significant uncertainty in the partial widths
occurs\footnote{Note that in the case of the three--body $H\to WW^*$ and $H\to
ZZ^*$ decay channels,  the virtuality of the off--shell gauge bosons is
important and the Higgs mass $M_H$, on which the partial decay widths depend
crucially as can be seen from Fig.~1, has to be very  precisely known.}.  

Doing so, one would have the following theoretically ``clean" observables
to consider: 
\beq
D_{ZZ} &=& \frac {\sigma( gg \to H \to ZZ)}{ \sigma( gg \to H \to VV)}
       = \frac {\Gamma( H \to ZZ)}{ \Gamma ( H \to VV)} \ \ = d_{ZZ} 
       \frac{c_Z^2} {c_V^2} \\
D_{WW} &=& \frac {\sigma( gg \to H \to WW)}{ \sigma( gg \to H \to VV)}
       = \frac {\Gamma( H \to WW)}{ \Gamma ( H \to VV)} =d_{WW} 
       \frac{c_W^2} {c_V^2} \\
D_{\tau\tau} &=& \frac {\sigma( gg \to H \to \tau\tau)}{ \sigma( gg \to H \to VV)}
       \ \ = \frac {\Gamma( H \to \tau\tau)}{ \Gamma ( H \to VV)} =\ d_{\tau\tau}
       \frac{c_{\tau}^2} {c_{V}^2} \\
D_{\gamma\gamma} &=& \frac {\sigma( gg \to H \to \gamma\gamma)}{ \sigma( gg 
\to H \to VV)}
       \ \ = \frac {\Gamma( H \to \gamma\gamma)}{ \Gamma ( H \to VV)} = 
       \ d_{\gamma\gamma}  
       \frac{c_{\gamma}^2}{c_{V}^2} 
\eeq
where  $c_{X}$ are the Higgs couplings to the $X$ states normalized to their SM
values\footnote{We will assume $c_X$ to be simply  (real) constants and do not 
consider anomalous vertices with, for instance, derivative Higgs couplings to
fermions or gauge bosons. The kinematics of the processes and the selection 
efficiencies are thus the  same as in the SM. Derivative (momentum--dependent)
Higgs couplings can be  checked, for instance, by evaluating the production
cross sections at  different c.m. energies  \cite{Rojo}.}, $c_X \equiv
g_{HXX}/g_{HXX}^{\rm SM}$, and the reduced decay  width ratios  $d_{XX}$, which
involve only gauge couplings and kinematical factors,  are displayed in Table 1
for the three possible normalisations.

\begin{table}[!h]
\renewcommand{\arraystretch}{1.3}
\begin{center}
\begin{tabular}{|c||c|c|c|c|c||c|c|}\hline
normalisation & $d_{WW}$ & $d_{ZZ}$ & $d_{\tau\tau}$ & $d_{bb}$ & $d_{\gamma\gamma}$ & 
$d_{\mu\mu}$ & $d_{\gamma Z}$   \\ \hline
$H\to ZZ$ & 8.14 & 1 & 2.39 & 21.9 & $8.64\times 10^{-2}$ & $8.33 \times 10^{-3}$ & 
$5.72 \times 10^{-2}$ \\ \hline
$H\to WW$ & 1 & 0.123 & 0.294 & 2.68 & $1.06\times 10^{-2}$ & $1.02 \times 10^{-3}$ & 
$0.72 \times 10^{-2}$ \\ \hline
$H\to VV$ & 0.89 & 0.11 & 0.261 & 2.39& $0.94\times 10^{-2}$ & $0.91\times 10^{-3}$ & 
$0.64 \times 10^{-2}$ \\ \hline
\end{tabular}
\end{center}
\vspace*{-0.5cm}
\caption{\small The reduced decay ratios $d_{XX}$ for the various final states 
$H\to XX$ observable at the LHC depending on the channel used as normalisation,
$H\to ZZ,WW$ or $VV=ZZ+WW$. These numbers are for $M_H=125$ GeV and are obtained
using the program {\tt HDECAY} \cite{HDECAY} for the Higgs branching ratios when
the SM inputs recommended   by the LHCHWG \cite{LHCXS} are used.} \label{dxx}
\vspace*{-0.2cm}
\end{table}

At this stage,  a few important remarks are in order.\vspace*{1mm} 

$i)$  To the observables of eqs.~(6--9),  one could add $D_{\mu\mu}$ and 
$D_{Z\gamma}$ for the channels $H\to \mu^+\mu^-$ and $H \to Z\gamma $  which
could be measured with some accuracy at the upgrade of the LHC when $\sqrt
s\approx 14$ TeV is reached and a large luminosity,  $\gsim 100$ fb$^{-1}$, is 
collected. The corresponding $d_{\mu\mu}$ and  $d_{\gamma Z}$ decay factors are
also given in Table 1 for completeness.\vspace*{1mm} 

$ii)$ For the loop induced $H\to \gamma \gamma$ channel (the situation is 
similar in the case of $H\to Z\gamma$), we have defined the reduced coupling
$c_{\gamma}$.   In the SM, this decay is mediated by loops involving mainly the
$W$ boson and the heavy top quark, with subleading contributions from the $b,c$
quarks and the $\tau$--lepton. In beyond the SM scenarios, not only the reduced
couplings  $c_{W}$ and $c_{f}$ are altered, but also new particles could
contribute to the loops. For $M_H \approx 125$ GeV and retaining only the
dominant  $W$ and $t$ contributions, the $c_{\gamma}$ coupling  where $\hat
c_{\gamma}$  represents the possible  contribution of new physics, can be 
written as    
\beq 
c_{\gamma} \approx  1.26 \times |c_{W}-0.21\, c_{t} + \hat c_{\gamma }| 
\eeq

$iii)$ We do not include the branching fractions for the $Z\to \ell^-  \ell^+$
and $W \to \ell \nu$ decays which are precisely known \cite{PDG}. In fact, in
the $H\to ZZ$ and $H\to WW$ decays, one could also include other channels such
as  $H\to ZZ\to \ell\ell \nu\bar \nu$ and $H\to WW\to \ell \nu jj$ if, in the
future, they turn out to be useful for a $\approx 125$ GeV Higgs boson.\vspace*{1mm} 

$iv)$ Finally, the decay ratios $D_{WW}$ or $D_{ZZ}$, depending on the chosen 
normalisation, are proportional to the ratio of squared couplings 
$c_{W}^2/c_{Z}^2$ and test custodial symmetry. They  are thus
related to the Veltman $\rho$ parameter \cite{Veltman} or, equivalently, to the
Peskin--Takeuchi $T$ \cite{Peskin} or Altarelli--Barbieri $\epsilon_1$ 
\cite{Altarelli} parameters,
\beq
c_{W}^2/c_{Z}^2 \approx  \rho \approx M_W^2/(\cos^2\theta_W M^2_Z) 
\approx 1+\alpha T \approx 1+\epsilon_1
\eeq
which from the high precision electroweak data has been shown to be very close
to unity\footnote{In addition, only very few new physics models (e.g.  models
with Higgs triplets  and  some composite models)  allow for deviation of this
ratio for unity at tree--level; for a  recent discussion, see
Ref.~\cite{custodial}. We also note that if custodial symmetry is violated,
the parameters $\rho$ etc... of eq.~(11) become sensitive to the ultraviolet cut--off 
and the equation becomes questionable.} \cite{PDG}.  
Assuming custodial symmetry, one could then assume $c_{W}=c_{Z}=c_{V}$ and  use
the combined $H\to WW$ and $H\to ZZ$ channels, 
\beq
\Gamma(H\to VV) =\Gamma(H\to WW)+ \Gamma(H\to ZZ)
\eeq
as a reference channel in eqs.~(6--9), to increase the statistical accuracy of
the normalisation   factor. The reduced decay values $d_{XX}$ in this case are
also given in Table 1.\smallskip  

The previous discussion assumes that one can consider only the dominant
$gg\to H$  production channel. Nevertheless, in practice, the other processes,  
in particular the vector boson fusion and the associated HV channels, 
contribute also to the total Higgs cross section and, more importantly, these
channels lead to specific  topologies that greatly facilitates the Higgs search
in some cases.  One should thus use a more accurate expression  compared
to eq.~(5) to define the decay ratios $D_{XX}$. Ignoring the $t \bar tH$
production channel for  the moment (but its small contribution can be readily 
included), one would then have  
\beq
D_{XX} &=& \frac {\epsilon_{gg}^X \sigma( gg \! \to\! H \!\to \!XX )\! + \!
\epsilon_{VBF}^X \sigma( qq \! \to \! Hqq \!\to \! qq XX ) \! + \! 
\epsilon_{HV}^X \sigma( q\bar q \! \to \! VH \! \to  \! VXX)}{\epsilon_{gg}^V 
\sigma( gg \! \to \! H \! \to \! VV )+ \epsilon_{VBF}^V
\sigma( qq \! \to \! Hqq \! \to \! qq VV ) +  \epsilon_{HV}^V \sigma( q\bar q 
\! \to \! VH \! \to VVV)} \nonumber \\
&=& \frac {\epsilon_{gg}^X \sigma( gg \! \to\! H)\! + \!
\epsilon_{VBF}^X \sigma( qq \! \to \! Hqq) \! + \! 
\epsilon_{HV}^X \sigma( q\bar q \! \to \! VH)}{\epsilon_{gg}^V 
\sigma( gg \! \to \! H)+ \epsilon_{VBF}^V
\sigma( qq \! \to \! Hqq) +  \epsilon_{HV}^V \sigma( q\bar q 
\! \to \! VH)} \times \frac {\Gamma( H \to XX)}{ \Gamma ( H \to VV)}
\eeq
where $\epsilon^X$ stands for the experimental efficiency to select the Higgs
events in the $gg$, VBF and HV channels. Note that we left the normalisation
channel $H\to VV$ unspecified in such a way that any of the $ZZ, WW$ or $
VV\!=\!WW\!+\!ZZ$ possibilities can be chosen. The second line of eq.~(13) shows
that still, some common systematical uncertainties and the uncertainties in the 
Higgs branching ratio cancel out. In addition, if the efficiencies $\epsilon^X$
and $\epsilon^V$ are comparable or one production channel is dominant, a large
part of the cross section uncertainties also  cancel out. There is thus still a
clear advantage in using these ratios.   

In fact, since the VBF and the HV channels involve  two additional jets (or leptons
in the case of HV) in the final state,  the $gg$ fusion mechanism can be singled
out by considering the  Higgs $+0\;$jet cross sections, i.e. by requiring that no 
hard jet (with $p_T^{\rm jet}$ larger than say 30--40 GeV) is produced along
with the Higgs particle. One can  have then almost pure $gg$ fusion events and
construct the ratio,  
\beq 
D_{XX}^{(0j)} = \frac {\sigma( gg \to H+0j \to XX)}{ \sigma( gg \to
H+0j \to VV)} = \frac {\Gamma( H \to XX)}{ \Gamma ( H \to VV)}  
\eeq 
As the additional jets are produced at higher orders in QCD, NLO or NNLO, the
Higgs+0jet cross sections represent a large fraction of the $gg\to H$ inclusive
rate: for $p_T^{\rm jet}\gsim 30$ GeV, one has {\it very crudely} $60\%, 30\%$ and 
10\% for, respectively, the $0, 1$ and $2\,$jet cross sections.  However, as was
recently realized, the breaking of the Higgs cross  sections into jet categories
introduces significant uncertainties 
\cite{Scale-H0j}. These additional uncertainties, if one adopts the same
criteria (same jet veto etc..)  for selecting the $XX$ and the $VV$ events in
the numerator and denominator of eq.~(14), will also cancel out in the ratio. 

Another remark is that one can also include to the $H+1j$ contributions in
eq.~(14), and  thus consider  the ratios $D^{(0+1j)}_{XX}$. This would increase
the number of signal events   without having too much
contamination from the VBF and VH processes. However, in the channel $H\to WW\to
\ell \ell  \nu \nu +1j$ for instance, one would have to deal with the large
$t\bar t \to b\bar b WW$ background in which one of the (untagged) $b$--jets is
soft and escapes detection.   

In turn, if one focuses on the VBF events which have  a special topology and, 
in most cases, have a more favorable signal to background ratio \cite{VBF}, 
one would consider the ratio
\beq
D_{XX}^{(2j)} &\hspace*{-2mm}= \hspace*{-2mm} & \frac {\epsilon_{gg}^X \sigma( gg \! \to\! Hjj \!\to \!XXjj
 )\! + \! \epsilon_{VBF}^X \sigma( qq \! \to \! Hqq \!\to \! qq XX ) \! + \! 
\epsilon_{HV}^X \sigma( q\bar q \! \to \! VH \! \to  \! q\bar q XX)}{\epsilon_{gg}^V 
\sigma( gg \! \to \! Hjj \! \to \! VVjj )+ \epsilon_{VBF}^V
\sigma( qq \! \to \! Hqq \! \to \! qq VV ) +  \epsilon_{HV}^V \sigma( q\bar q 
\! \to \! VH \! \to q\bar q VV)} \hspace*{-5mm} \nonumber \\
&\hspace*{-2mm}=\hspace*{-2mm}& \frac {\epsilon_{gg}^X \sigma( gg \! \to\! Hjj)\! + \!
\epsilon_{VBF}^X \sigma( qq \! \to \! Hqq) \! + \! 
\epsilon_{HV}^X \sigma( q\bar q \! \to \! VH)}{\epsilon_{gg}^V 
\sigma( gg \! \to \! Hjj)+ \epsilon_{VBF}^V
\sigma( qq \! \to \! Hqq) +  \epsilon_{HV}^V \sigma( q\bar q 
\! \to \! VH)} \times \frac {\Gamma( H \to XX)}{ \Gamma ( H \to VV)}
\eeq
While most of the events from the HV process can be removed by requiring that 
the jet-jet invariant mass does not coincide with $M_W$ or $M_Z$, a significant
fraction of the $gg\! \to\! H\!+\!2j$ events (of the order of 30\% \cite{Hjj})
will remain even  after the specific cuts that select the VBF configuration are
applied. Hence, despite the very small scale and PDF uncertainty that affects
the inclusive VBF Higgs cross section \cite{LHCXS}, the contamination by the 
$gg\! \to\! H\!+\!jj$ channel in which the combined scale+PDF uncertainty is  at
the level of $\approx 50\%$,  will make the total  uncertainty  in the
$H\!+\!jj$ final sample very large. Again, by performing the ratio of eq.~(15),
one could reduce, if not almost completely eliminate, the theoretical
uncertainty in the  extraction of the Higgs couplings from these processes. 

One can also use the same procedure in the case of the Higgs+1jet configuration
which, for instance, can be appropriate in the $H\to \tau \tau$ search  channel
\cite{Bruce}. In fact, this configuration would be extremely  useful in the case
of invisible Higgs decays which can be searched for  in monojet events in the
process $gg\!\to\! H\!+\!1j$ with $H\to$ invisible as has been recently
discussed in Ref.~\cite{portal} for instance. In this case, one could consider 
the  ratio, 
\beq
D_{\rm inv}^{(1j)} = \frac {\sigma( gg \to H+1j \to 1j+E_T\hspace*{-3mm}\slash 
)}{ \sigma( gg \to H+1j \to 1j+VV)}
       = \frac {\Gamma( H \to {\rm inv})}{ \Gamma ( H \to VV)}        
\eeq 
in which, again, the theoretical uncertainties in the $gg\to H$ cross section,  
which could mimic the additional invisible contribution to the total Higgs
decay width, will cancel out.

Of course, ultimately, the ratios $D_{XX}$ for a given final state and from 
different jet configurations should be combined to reach a better statistical 
accuracy.

Let us now turn to the $H\to b\bar b$ final state which deserves a special 
treatment as it is observable mainly (if not exclusively) in the  $q \bar q\! 
\to \! HV \!\to\!  b\bar b V$ process using boosted jet techniques to isolate 
the $b\bar b$ events \cite{Gavin}. One can use the process as it is to measure 
BR($H\to b\bar b)$ as the cross section $\sigma (q\bar q \to HV)$ is  predicted
with an accuracy of $\approx 5\%$ \cite{LHCXS} that will be much smaller than
the experimental error. But one can also  consider the ratio  
\beq 
D_{bb} = \frac {\sigma( q\bar q \to HV \to b\bar b V)}{ \sigma( q \bar q  \to
HV  \to  VVV)} = \frac {\Gamma( H \to b\bar b )}{ \Gamma ( H \to VV)} = d_{bb}
\frac{c_b^2}{c_V^2} 
\eeq  
However, as the cross sections times branching ratios for the clean $H\! \to\!
ZZ \! \to \! 4\ell $ or  $H\!\to\! WW\! \to\! \ell \ell \nu \nu$ final states
are very small, the normalisation above might not be appropriate. 

If the Higgs signal could be extracted in the $q\bar q \to HV \to \tau^+\tau^-
V$ channel (as, for instance,  recently advocated in Ref.~\cite{tau-new}), the
situation  would be rather straightforward as one could  simply consider the
ratio of $b\bar b V$ to $\tau^+\tau^- V$ production, $D_{bb/\tau\tau} = \sigma(
q\bar q \to HV \to b\bar b V)/ \sigma( q \bar q  \to HV  \to  V\tau\tau) =
\Gamma( H \to b\bar b )/ \Gamma ( H \to \tau\tau )$  which directly provides
the  important ratio $c_b^2/c_\tau^2$, that allows to test the hierarchy of the
Higgs--fermion couplings and the important SM prediction, $c_b^2/c_\tau^2
\approx 3 \bar m_b(M_H^2)/m_\tau^2  \approx 10$.

\subsection*{3. The cross section ratios $\mathbf{C_{XX}}$}

So far, we have only considered a given production process with different decay
channels and constructed decay ratios $D_{XX}$ in which the theoretical 
uncertainties in the cross sections as well as some model dependence due to the
Higgs total decay width should cancel out. However, when doing so, some very
important information that is contained in the cross section only has been
removed. This was, for instance, the case of the Higgs to gluons coupling which
generates  the $gg\to H$ process. In this  section, we  briefly consider ratios
of cross sections for different production processes but for a given Higgs decay
channel in which  this information is retained. In these cross section ratios,
that we denote by $C_{XX}$, it is the branching fractions or the partial decay
width (and hence the ambiguities in the total width)  which will cancel out in
addition to some systematical errors that are common to the two processes,
leaving us only with the theoretical uncertainties due to the  production cross
sections. In the absence of cleaner ratios to probe the couplings involved in
the cross sections in an almost model  independent way, and in order not to
loose the crucial information that they provide, this choice can be considered 
as a ``lesser evil".

We start by reconsidering the determination of the $Hbb$ coupling from the HV
process. Instead  of performing the decay ratio of eq.~(17) in which  the
normalisation might not be appropriate, one can take advantage of the fact that
the HV and the VBF (inclusive) rates are affected  by rather small theoretical
uncertainties and consider the cross section ratio 
\beq 
C_{bb} = \frac {\sigma( q\bar q \to HV \to b\bar b V)}{ \sigma( q q  \to Hqq
\to  VVqq)} \propto \frac {\Gamma( H \to b\bar b)}{ \Gamma ( H \to VV)} \propto 
\frac{c_b^2}{c_V^2} 
\eeq  
As the production cross sections in both processes are proportional to the 
square of the coupling $c_{V}$, it will cancel out in the ratio\footnote{Note
that in $\sigma(pp \to HZ)$, there is a contribution  from the $gg\to HZ$ box
diagram which is not proportional to $c_V^2$ \cite{ggHZ}  and which is about
5--10\% depending  on the considered c.m. energy.} leaving only the dependence
on the decay ratios (this is particularly the case if one assumes the custodial
symmetry which enforces the relation $c_{W}=c_{Z}$ that simplifies the problem).
Of course, different systematics will enter the two processes and they have to
be taken care of. However, at   least  the uncertainties from the luminosity,
the Higgs total width and eventually also part of the uncertainty due to the
PDFs (as both processes are initiated  by incoming quarks) will cancel out.
The usual systematical experimental errors in the selection of the 
two channels as well as the very small scale uncertainty  that affect the two
processes remain though. This nice picture is, however, spoilt 
by the contamination of the VBF process by the $gg\!\to\! Hjj$ contribution. 

Another  issue is  the determination of the important Higgs coupling to
top quarks. This coupling can be first determined indirectly from the $gg\to H$
cross section which, as discussed earlier, is dominantly generated by a top  
quark loop  and hence is proportional to $g_{Htt}^2$.   By normalizing
to the VBF process, one would  have the ratio,  
\beq
C_{gg} = \frac {\sigma( gg \to H \to VV)}{ \sigma( q q  \to Hqq \to 
VVqq)}
       \propto  \frac {c_g^2}{c_V^2}
\eeq 
which nevertheless includes the large theoretical uncertainty that affects  the
$gg\!\to\! H$ rate and, eventually,  the comparable one of VBF  if the
contamination  by $gg\to Hjj$ events remains large.  One can chose a decay
normalisation with  $V\!=\!W,Z$ or $W\!+\!Z$ but one could also add the case 
$V\!=\!\gamma$ to increase the statistics; the $H\gamma\gamma$ coupling, which
is also very sensitive to new  physics, will anyway drop in the ratio. The
coupling $c_g$ receives a dominant contribution from the top quark, but also a
smaller one from the bottom quark; contributions from new strongly interacting
particles are also possible (see also Ref.~\cite{Chris}):   
\beq 
c_g \approx  1.075 \times |c_{t}-(0.066 +0.093\,i)\; c_{b} + \hat c_{g}| 
\eeq

Nevertheless, one should consider the previous ratio as a measurement of the
Higgs to gluons coupling, rather than the $Ht\bar t$ coupling, as there is the
possibility of  loop contributions from new strongly interacting particles that
couple to the Higgs boson.   

A more direct measurement of the $H t\bar t$
coupling can be performed  in the $pp \to t\bar tH$ process,  once enough data
is collected.  In  this case, one could consider the ratio\footnote{  Note that
in the $ pp \to t\bar tH$ process, the PDF uncertainty is the largest source of
error and is about $\pm 10\%$ for $M_H=125$ GeV at $\sqrt s=14$ TeV
\cite{LHCXS}. One could reduce this uncertainty by normalizing the cross section
to the $pp\to t\bar t$ rate,  eventually at high enough invariant $t\bar t$ mass
to be in the same kinematical regime as in the $pp \to t\bar tH$ process. We
thank R. Godbole for a discussion on this point.} $C_{tt} = \sigma( pp \to H
t\bar t \to t\bar t VV)/ \sigma( q q  \to Hqq \to VVqq)$ which is proportional
to $c_t^2/c_V^2$. However, as the most interesting process in this context 
(if it is made viable experimentally) is 
$pp \to H t\bar t$ with $H\to b\bar b$, the proper normalisation to use should
be the $q \bar q   \to VH  \to V b\bar b$ channel
\beq
C_{tt} = \frac {\sigma( pp \to H t\bar t \to t\bar t b\bar b)}{ \sigma( q \bar 
q  \to VH  \to V b\bar b)}
       \propto \frac {c_t^2}{c_V^2}
\eeq 
In both cases eqs.~(19) and (21),  the branching fractions and some common 
systematical errors have cancelled out, leaving us only with the theoretical
uncertainties due to the $gg\to H$ and $q\bar q/gg\to  t\bar t H$ production
cross sections, as the one affecting at least the (inclusive) HV channel
is particularly small. 

A final word is due to the Higgs self--coupling which, at the LHC, can be only
determined from double Higgs production in the $gg\to HH$ process once a very  
high luminosity is collected \cite{HHH}. As the process is initiated by gluons 
similarly  to the $gg \to H$ case, and the NLO QCD corrections in  both processes
are very  similar \cite{DDS}, a large component of the QCD uncertainties should
drop if one considers the ratio
\beq
C_{HH} = \frac {\sigma( gg \to HH)}{ \sigma( gg \to H)}
       \propto (a g_{HHH}+ b g_{Htt})^2 \times \frac{  
       {\rm BR}(H\to XX)\cdot {\rm BR}(H\to YY) } { {\rm BR}(H\to XX)}
\eeq 
and one would be mostly left with the smaller branching fractions uncertainties.

\subsection*{4. Application of the ratios to the  LHC data}

It is clear that a truly reliable estimate of the experimental accuracies in
the   determination of the ratios of cross sections discussed previously can
only come from the  ATLAS and CMS collaborations as they have the full
information on the systematical errors that affect their measurements and the
experimental efficiencies to select the various observed channels.
Nevertheless,  in order to illustrate the usefulness of the ratios that we have
introduced in this paper, we will attempt in this section to provide a rough
estimate of the accuracies that can be obtained on some of these ratios, using
the partial information that was provided in the combined ATLAS \cite{ATLAS} and
CMS \cite{CMS} analyses. 

The ATLAS collaboration has
given the best fit values and  the corresponding  uncertainties of the signal
strength $\hat \mu$ in the inclusive search channels $H\to WW^*\to \ell \ell \nu
\nu, ZZ^*\to 4\ell$ and $\gamma\gamma$  when the values at $\sqrt s=7$ and 8 TeV
are combined for $M_H=126$ GeV  (Table 7 of Ref.~\cite{ATLAS}); these $\hat \mu$
values are listed in Table 2. Instead, the CMS collaboration did not provide the
exact $\hat \mu$ values in these channels but reported them in a  figure for
$M_H=125.5$ GeV (Figure 19 of Ref.~\cite{CMS}); the numbers listed  in Table 2
are thus approximate. The exact values of $\hat \mu_{\gamma\gamma jj}$  for the
$H\to \gamma \gamma$ channel in the VBF configuration  have  not be provided by
both  collaborations  and the approximate one listed for ATLAS in Table 2 is
taken from   Fig.~14 of Ref.~\cite{ATLAS-gg}. We do not consider the $H\to
\tau\tau$ and $H\to b\bar b$ channels as, with the present data,  the
uncertainties are too still large.  

\begin{table}[!h]
\renewcommand{\arraystretch}{1.3}
\begin{center}
\begin{tabular}{|c||c|c|c||c|}\hline
& $\hat \mu_{WW}$ & $\hat \mu_{ZZ}$ & $\hat \mu_{\gamma\gamma}$ & 
$\hat \mu_{\gamma \gamma jj}$   \\ \hline
ATLAS & $1.3 \pm 0.5$ & $1.4 \pm 0.6$ & $1.8 \pm 0.5$ & $2.7 \pm 1.3$ \\ \hline
CMS & $0.67 \pm 0.4$ & $0.72 \pm 0.5$ & $1.6 \pm 0.5$ & -- \\ \hline
\end{tabular}
\end{center}
\vspace*{-0.3cm}
\caption{\small The ATLAS and CMS signal strength  modifiers $\hat \mu$  in the
various search channels that are used for our illustration; they are obtained 
from Refs.~\cite{ATLAS,CMS,ATLAS-gg}.}  
\label{hatmu}
\vspace*{-0.2cm}
\end{table}

We will first construct the cross section ratios in the inclusive search
channels. For this purpose, we will make the following assumptions which seem 
to us rather reasonable:  $i)$ the efficiencies for selecting the $H\to
WW,ZZ,\gamma \gamma$ modes  are approximately the same so that the cross section
part in eq.~(13) drops out and we are left only with the ratios of partial decay
widths; $ii)$ the uncertainties in the measurements are dominated by the
statistical error as well as  by systematical errors that are
assumed  to be uncorrelated in the different channels, can be treated as
Gaussian (the theoretical and the  common systematical errors from the
luminosity and the Higgs branching ratios will drop in the ratios); $iii)$ 
the remaining 
uncertainties in the ATLAS and CMS results are assumed to be uncorrelated and
the results of the two experiments can be averaged.

With these assumptions, we first construct the ratios $D_{ZZ}$ and $D_{WW}$,
first for the individual experiments and, then, when the ATLAS and CMS results
are averaged; to increase the statistics we use the normalisation in which the 
channels $H\to ZZ$ and $H \to WW$ are combined. We obtain in this case, when 
averaging the ATLAS and CMS results 
(from now on and for simplicity, we will set the factors $d_{XX}$ given in 
Table 1 to unity) 
\beq 
D_{WW} &\equiv & c_W^2/c_V^2= 0.97 \pm 0.40 \nonumber \\ 
D_{ZZ} &\equiv & c_Z^2/c_V^2= 1.04\; \pm 0.46 
\eeq 
where the errors are only of experimental nature and mostly statistical. 
Remarkably, these values are already close to unity with the present data, 
showing that custodial symmetry approximately holds\footnote{Because we are
combining the $H\!\to \! WW$ and $H\! \to \! ZZ$ channels for the normalisation
and the ATLAS and  CMS $\hat \mu_{WW}$ and $\hat \mu_{ZZ}$ values, our result is
more accurate than the one given by the CMS collaboration,  $c_{W}^2/c_{Z}^2
\equiv R_{WW/ZZ}=0.9^{+1.1}_{-0.6}$ ~\cite{LHC}.}.    At the end of this year,
when $\approx 30$ fb$^{-1}$ of data will hopefully be collected  by both the ATLAS and CMS
experiments, custodial symmetry can be checked at the $\approx 25\%$ level.
Ultimately, if more than 300 fb$^{-1}$ of data is collected by the ATLAS and CMS
collaborations at $\sqrt s \approx 14$ TeV, these relations  can be checked at
the $\approx 5\%$ level, with no limitation from theoretical uncertainties and
hopefully also from  systematical uncertainties if the two channels $H\to
WW,ZZ$ are analyzed  in the same way.     

A second and extremely important ratio  which can be already constructed from
the ATLAS and CMS signal strength modifiers in the inclusive channels is 
$D_{\gamma \gamma}$. One obtains, again by combining the ATLAS and CMS results
and using  the combined  $H\to VV=WW+ZZ$ channel as a normalisation,   
\beq 
D_{\gamma\gamma} &\equiv & c_\gamma^2/c_V^2= 1.70 \pm 0.43   
\eeq   
which is in accord with the SM expectation at the 95\% confidence level (CL).  
Again,  this ratio should be  free of theoretical and common systematical
uncertainties  and could be probed with an accuracy at the level of 25\% at the
end of this year  and, ultimately, at the 5\% level at the upgraded LHC. We
expect this  measurement to be the most important one to be performed in the
Higgs sector at the LHC. It is  crucial because first, it might involve
contributions from new light charged particles that couple to the Higgs boson
and second, because it measures  the relative strength of the Higgs couplings to
vector bosons and to the heavy top quark, eq.~(10), which is one of the most
important checks of the Weinberg generalisation \cite{Weinberg} of the
Englert--Brout--Higgs mechanism to fermions.

The left--hand panel of Fig. 2 shows the 68\% CL contour in the $[c_W, c_t]$
plane that is allowed by the present data assuming custodial symmetry. Shown
also are the contours that can be probed at $\sqrt s=8$ TeV  with an integrated
luminosity of 30 fb$^{-1}$ per experiment  and at $\sqrt s= 14$ TeV with 300 
fb$^{-1}$ data, assuming again that  the measurement will only be limited by
statistics  and that the central value stays the same as presently.  The few
percent accuracy on the relative Higgs couplings $c_W$ and $c_t$ that can be
achieved at the end of the planed LHC run is not only due to the increase of the
luminosity to $300$ fb$^{-1}$ but, also, to the increase of the $gg\!\to \!H$ 
inclusive cross section by a factor $\approx 2.5$ when  increasing the  energy 
from $8$ TeV to   $14$ TeV. Note that in the absence of new physics
contributions $\hat c_\gamma$ to the $H\gamma\gamma$ coupling as  is assumed in
Fig.~2, the $c_t$ values which  best fit the data are negative  to accommodate
the presently significant deviation of  $D_{\gamma\gamma}$ from unity.

\begin{figure}[hbtp]
\begin{center}
\vspace*{-.1cm}
\epsfig{file=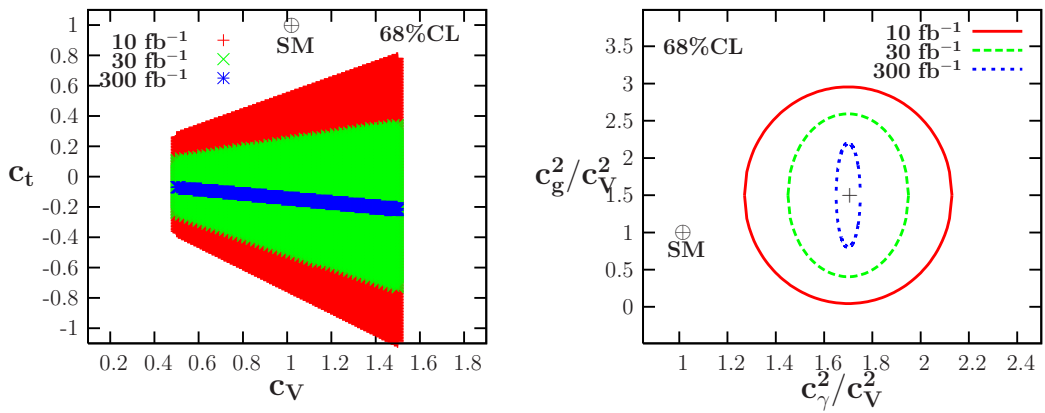,width=15cm}
\vspace*{-.2cm}
\caption[]{\small The 68\% CL contours for the couplings $c_W,c_t$ as presently
allowed from the measurement of the ratio $D_{\gamma\gamma}$ at  the LHC (left)
and the 68\% CL allowed range in the plane $[C_{gg}^{-1}, D_{\gamma  \gamma}]$ with
and without the theoretical uncertainty (right). The prospects for $\sqrt  s=8$
TeV and 30 fb$^{-1}$ data and $\sqrt s=14$ TeV and 300 fb$^{-1}$ data are also
displayed  in both cases.}
\end{center}
\label{fits}
\vspace*{-.5cm}
\end{figure}

The other  measurement which could be already exploited is the  cross section
for the VBF--like associated production of $\gamma\gamma$  events with dijets.
In this case, the exact values of the strength modifier $\hat \mu_{\gamma\gamma
jj}$ have not been given by both collaborations\footnote{ In Ref.~\cite{CMS-gg},
the CMS collaboration provided some numbers (again in a  plot) for two
configurations of the $\gamma\gamma$--dijet sample, with loose and  tight cuts
on the di--jets, and separately for $\sqrt s=7$ and 8 TeV.  Since we do not have
all the information, we refrain from combining the numbers to obtain a global 
value for the strength modifier in this case. We emphasize again the fact that
these numbers are very important (in contrast to the ones for the boosted 
decision trees which are irrelevant for our purpose) and should be part of the
public information that is provided by the ATLAS and CMS experiments.}   and the
approximate  one that we quote for ATLAS in Table 2 is  obtained from  Fig.~14
of Ref.~\cite{ATLAS-gg} with $M_H=126.5$ GeV. Besides the experimental
uncertainty of $50\%$, we will assume an additional theoretical uncertainty of
$\approx 20\%$ which is mainly due to the contamination of the VBF channel by
the $\approx 30\%$ $gg \to H+jj$ contribution     which is affected by a
$\approx 50\%$ uncertainty as was discussed previously.  This gives  a value
$\hat \mu_{\gamma\gamma jj}=2.7 \pm 1.3 \pm 0.5$ in which the  two uncertainties
are to be added linearly since the theoretical uncertainty has no statistical
ground and should be considered as a bias rather than  a mere nuisance.

Nevertheless, one cannot use the full potential of this measurement for the time
being as there is  no precise result for a corresponding measurement of the VBF
$H\to WW,ZZ$ or $\tau\tau$ channels  to perform the ratios of eq.~(15) in which
the large  theoretical uncertainties cancel out and which,  for instance, could
be combined with $D_{\gamma \gamma}$ in eq.~(24) to reach a better level of accuracy. 
Hence, one can use $\hat \mu_{\gamma\gamma jj}$ as it is, but we prefer again to
construct a ratio, the one of eq.~(19), to eliminate possible ambiguities  from
the Higgs branching fractions or the total Higgs decay width.  The price to pay
is the introduction of the large  theoretical uncertainty that affects the
inclusive $gg\to H$ cross section, that we take to be $\pm 20\%$, and which need
to be added to the one affecting the VBF--like cross section. This will allow to
determine in a non ambiguous way the ratio of couplings $c_{g}^2/c_V^2$, which
provides additional interesting information. One obtains using the ATLAS result
only
\beq 
C_{gg}^{-1}= \frac{\hat \mu_{\gamma\gamma jj} } {\hat \mu_{\gamma\gamma } } 
= c_V^2/c_{g}^2= 1.5 \times [1 \pm 0.57 \pm 0.4] 
\eeq
where the first uncertainty is experimental and the second one theoretical. 
This ratio is confronted to $D_{\gamma\gamma}$ in the right-hand side of Fig.~2
where the 68\% CL contour $[C_{gg}^{-1}, D_{\gamma \gamma}]$ is displayed for
the  present data, as well as for the projections for the end of this year and
the end of the LHC run. One can see that while the error on $D_{\gamma \gamma}$
shrinks  considerably when more data is collected, as it is mainly  due to
statistics, the total error  on $C_{gg}^{-1}$  will be limited to the 40\%
theoretical uncertainty  which can be reduced only  with a more refined
calculation of the $gg\to H$ cross section in the various jet categories.

Instead, when more data will be collected by ATLAS and CMS, other Higgs decay 
channels  could be probed in the VBF configuration and ratios such as
$D_{\gamma\gamma}^{(jj)}$ and $D_{\tau\tau}^{(jj)}$ could be then determined and
would allow a very precise measurement of  ratios of Higgs couplings in a way
that is complementary to what is obtained in the inclusive $gg\to H$
mode. The power of the VBF mechanism could be then fully exploited.

\subsection*{5. Conclusion}

We suggest to use ratios of Higgs production cross sections  at the LHC for
different Higgs decay channels  such as $H\to WW,ZZ,\tau\tau, b \bar b, \gamma
\gamma$ and eventually $H\to \mu^+\mu^-, Z\gamma $, to determine the Higgs
couplings to fermion and gauge bosons in a way that is not limited by
theoretical uncertainties. These uncertainties, which are large being of order
$\approx 20\%$, affect not only the main production channel, $gg\to H$, but also
the vector boson fusion channel $qq \to Hqq$ as it is  significantly
contaminated by the $gg\to Hjj$ contribution. The observables $D_{XX}$ that we
propose  involve ratios of Higgs partial decay widths and are hence also free
from some systematical errors, such as the one from the luminosity measurement,
and from other theoretical ambiguities such as those involved in the Higgs
branching ratios or total decay width. In this respect, there is less model
dependence in these ratios when  beyond the SM scenarios are considered since
they do not involve the Higgs total width. 

One can also construct ratios of cross sections for different production
processes with a given Higgs decay, $C_{XX}$,  in which some ambiguities drop
out, but  these are less powerful than the ratios $D_{XX}$ as the theoretical
uncertainties that affect the cross sections remain. 

A rough analysis with the $\approx 10~{\rm fb}^{-1}$ data collected by ATLAS and
CMS shows that some of these ratios are compatible with the SM expectation.  At
the end of the LHC run with $\sqrt s=14$ TeV and $\approx 300~{\rm fb}^{-1}$
data per experiment, some ratios can be determined with a very high  accuracy, 
at the 5\% level, without any limitation from theoretical
uncertainties.

Hence, the LHC could become a precision machine for Higgs physics provided that
ratios of cross sections times branching fractions for the same production
channel, with eventually the same selection cuts for the different final state
topologies,  are considered. \bigskip

\noindent {\bf Acknowledgements}: I thank the CERN Theory Unit  for  hospitality
and J. Baglio, A. Falkowski, R. Godbole, C. Grojean,    B. Mellado and D.
Zeppenfeld for discussions. 

\end{document}